\documentclass[twocolumn,secnumarabic,amssymb, nobibnotes, aps, prc, superscriptaddress, nobalancelastpage,longbibliography,nofootinbib]{revtex4-1}

\usepackage{epsfig}
\usepackage{graphicx}
\usepackage{stmaryrd}
\usepackage{amssymb,tabularx,dcolumn}
\usepackage{supertabular,ltxtable}
\usepackage{longtable}
\usepackage{amsmath}
\usepackage{amstext}
\usepackage{float}
\usepackage{color}
\usepackage{bm}
\usepackage{hyperref}
\hypersetup{breaklinks=true,colorlinks=true,linkcolor=blue,citecolor=blue,filecolor=magenta,urlcolor=cyan}
\usepackage{mathtools}
\usepackage{multirow}
\usepackage{makecell}
\usepackage[utf8]{inputenc}
\usepackage{tabu}
\usepackage{braket}
\usepackage{subfigure}
\usepackage{enumitem}

\usepackage{xr-hyper}
\usepackage{hyperref}
\hypersetup{breaklinks=true,colorlinks=true,linkcolor=blue,citecolor=blue,filecolor=magenta,urlcolor=cyan}

\usepackage[all]{hypcap}

\usepackage{xcolor}
\definecolor{pastelgray}{rgb}{0.81, 0.81, 0.77}
\definecolor{beaublue}{rgb}{0.9, 0.9, 0.93}

\makeatletter
\def\@bibdataout@aps{%
\immediate\write\@bibdataout{%
@CONTROL{%
apsrev41Control%
\longbibliography@sw{%
    ,author="08",editor="1",pages="1",title="0",year="1"%
    }{%
    ,author="08",editor="1",pages="1",title="",year="1"%
    }%
  }%
}%
\if@filesw \immediate \write \@auxout {\string \citation {apsrev41Control}}\fi
}
\makeatother

\newcommand{\SV}{SV-min}
\newcommand{\Fy}{Fy($\Delta r$,BCS)}

\newcommand{\Rch}{$R_{\rm ch}$}

\begin{document}

\title{Statistical correlations of nuclear quadrupole deformations and charge radii}

\author{Paul-Gerhard Reinhard}
\affiliation{Institut für Theoretische Physik, Universität Erlangen, Erlangen, Germany}

\author{Witold Nazarewicz}
\affiliation{Facility for Rare Isotope Beams and Department of Physics and Astronomy, Michigan State University, East Lansing, Michigan 48824, USA}

\date{\today}

\begin{abstract}
\begin{description}
\item[Background]
 Shape deformations and charge radii, basic properties of atomic nuclei, are influenced by both the global features of the nuclear force and the nucleonic  shell structure. As functions of proton and neutron number, both quantities  show regular patterns and, for nuclei away from magic numbers, they change very smoothly from nucleus to nucleus.

\item[Purpose]
In this paper, we explain how the local shell effects are impacting the statistical correlations between quadrupole deformations and charge radii in well-deformed  even-even Er, Yb, and Hf isotopes. This implies, in turn, that sudden changes in correlations can be useful indicators of underlying shell effects.

\item[Methods]
Our theoretical analysis  is performed in the framework of self-consistent mean-field theory  using  quantified  energy density functionals and density-dependent pairing forces. The statistical analysis is carried out
by means of the  linear least-square regression.

\item[Results]
The local variations of nuclear quadrupole deformations and charge radii, explained in terms of occupations individual deformed Hartree-Fock orbits, make and imprint on statistical correlations of computed observables. While the calculated deformations or charge radii are, in some cases, correlated with those of their even-even neighbors, the correlations seem to deteriorate rapidly with particle number.

\item[Conclusions]
The statistical correlations between nuclear deformations and charge radii of different nuclei are affected by the underlying shell structure. Even for well deformed and superfluid nuclei for which these observables change smoothly, the correlation range usually does not exceed $\Delta N=4$ and $\Delta Z=4$, i.e., it is rather short.
This result suggests that the frequently made assumption of 
reduced statistical errors for the differences between
smoothly-varying observables cannot be generally justified.

\end{description}
\end{abstract}
\maketitle

\section{Introduction}

The global behavior of nuclear radii and quadrupole deformations
is impacted by  the macroscopic properties of nuclear liquid drop such as incompressibility or surface tension. The local behavior, on the other hand, is determined by the microscopic quantal effects such as the nucleonic shell structure, nucleonic pairing, and zero-point correlations due to the particle-vibrational coupling. 

The origin of nuclear deformations can be traced back to the nuclear  Jahn-Teller effect \cite{Rei84e,Nazarewicz1993,Nazarewicz1994}, i.e., the spontaneous symmetry breaking of  the internal density  (or mean-field) due to the coupling of degenerate nucleonic states with collective surface vibrations of the nucleus.
The systematic behavior of nuclear quadrupole deformations can be explained
through the geometrical properties, or shell topology,  of valence proton and neutron orbitals \cite{Bertsch1968,Rei84e}.
According to the 
Hartree-Fock (HF) analysis  \cite{Dobaczewski1998,Werner1994}, the main contribution to the quadrupole deformation energy comes from the effective neutron-proton  quadrupole interaction that maximizes around the middle of proton and neutron shells. This results in 
simple patterns \cite{Janecke1981,NazRag1996} of quadrupole deformations
which can be well systematized in terms of the promiscuity factor \cite{Casten1987} that depends on 
the distance of $Z$ and $N$ to the closest magic proton and neutron
number. Atop this general behavior, local fluctuations in quadrupole deformations may occur due to occupations of individual deformed Nilsson single-particle (s.p.) orbits close to the Fermi level. Depending on their s.p. quadrupole moments, these orbits can increase or decrease total quadrupole deformations by polarizing the system.

Similar considerations pertain to nuclear charge radii, which are the monopole moments of the nuclear charge density that is dominated by the proton  density. Here, the occupations of states with large oscillator quantum numbers  dominate the general pattern. The charge radii are also impacted by nuclear deformations in the second order.

The purpose of this Paper is to analyse the local trends of quadrupole deformations and charge radii in terms of statistical correlations between predicted observables in neighboring nuclei. The motivation behind the use of statistical correlation approach can be explained as follows. If the  occupations of s.p.levels change smoothly with particle number, and the character of s.p.levels around the Fermi level is similar, one would expect to see large statistical correlations
between deformations and radii in close-lying  isotopes and isotones. On the other hand, if the intrinsic structure changes rapidly due to, e.g., crossings of s.p. levels with very different quantum numbers, the statistical correlations are expected to be reduced. The isotopic and isotonic trend of statistical correlations can thus be a useful guide in several respects. It indicates changes in shell structure important for model understanding and development. And the information about the typical  range of statistical correlation between nuclear observables is important for modeling  emulators based on machine learning \cite{Neufcourt2018,Neufcourt2020a} and assessing statistical errors on differences of observables, e.g.,
energy differences or differential radii.

This Paper is organized as follows. Section~\ref{theory} describes theoretical models used and the statistical correlation analysis. The results obtained in this study are presented in Sec.~\ref{results}. Finally, the conclusions are presented in Sec.~\ref{conclusions}.

\section{Theoretical models}\label{theory}
Our analysis has been carried out within the  self-consistent  nuclear energy density functional method \cite{Bender2003}. 
In our applications, we employ the energy density functionals (EDFs) {\SV} \cite{Klupfel2009} and {\Fy}  \cite{Reinhard2017a}. Both 
 have been optimized to large experimental calibration datasets of nuclear ground state data  by means of the standard linear regression technique, which provides
information on  uncertainties and statistical correlations
between observables.  

In {\SV} calculations, we employed the standard \cite{Reinhard2017a} density-dependent pairing force of mixed type \cite{Dobaczewski2001}. The generalized pairing functional in the {\Fy} model additionally depends on the gradient of nucleonic density
\cite{Fayans2000}. In both variants, pairing is treated in the BCS  approximation.  
We do that because the nuclei of interest are well bound; hence, the HF+BCS approach is expected to offer a reasonable description of pairing and high accuracy when computing statistical covariances.  Thus we employ {\Fy} as a BCS-analogue of the original {Fy($\Delta r$,HFB)} tuned to exactly 
the same calibration dataset, particularly the differential charge radii.

Our statistical correlation analysis  is based on linear least square regression \cite{Klupfel2009,Dob14a} using the covariance matrices obtained in the course of EDF calibration.
The correlation between quantities $x$ and $y$ can be quantified in terms of the bivariate correlation coefficient
\begin{equation}\label{COD}
R_{x,y}=\frac{\mathrm{cov}(x,y)}{\sigma_x\sigma_y},
\end{equation}
where  $\sigma_x$ and $\sigma_y$ are variances of $x$ and $y$, respectively.
The  square $R^2$ is the coefficient of determination (CoD) \cite{Glantz}. 
It contains information on how well one quantity is determined by another one, within a given model. For our earlier applications of CoD to nuclear observables, see
Refs.~\cite{Erler2015,Reinhard16,Schuetrumpf17,Reinhard2018d,Reinhard2020}.
Values of
CoD range from 0 to 1, where 0 implies that, for a given model,  the quantities $x$ and $y$  are
 uncorrelated, whilst 1 denotes that one quantity determines the
other completely.  

The correlation coefficient (\ref{COD}) is useful when estimating the variance of a difference $x-y$:
\begin{equation}
\sigma^2_{x-y}=\sigma^2_x+\sigma^2_y-2 R_{x,y}\sigma_x    \sigma_y.
\end{equation}
In particular, if the observables $x$ and $y$ are very well correlated
($R_{x,y}\approx 1$),  the variance of a difference becomes
$\sigma_{x-y}\approx |\sigma_x-\sigma_y|$, i.e., it can be very small if $ \sigma_x \approx \sigma_y$.

\begin{figure}[!htb]
\includegraphics[width=1.0\columnwidth]{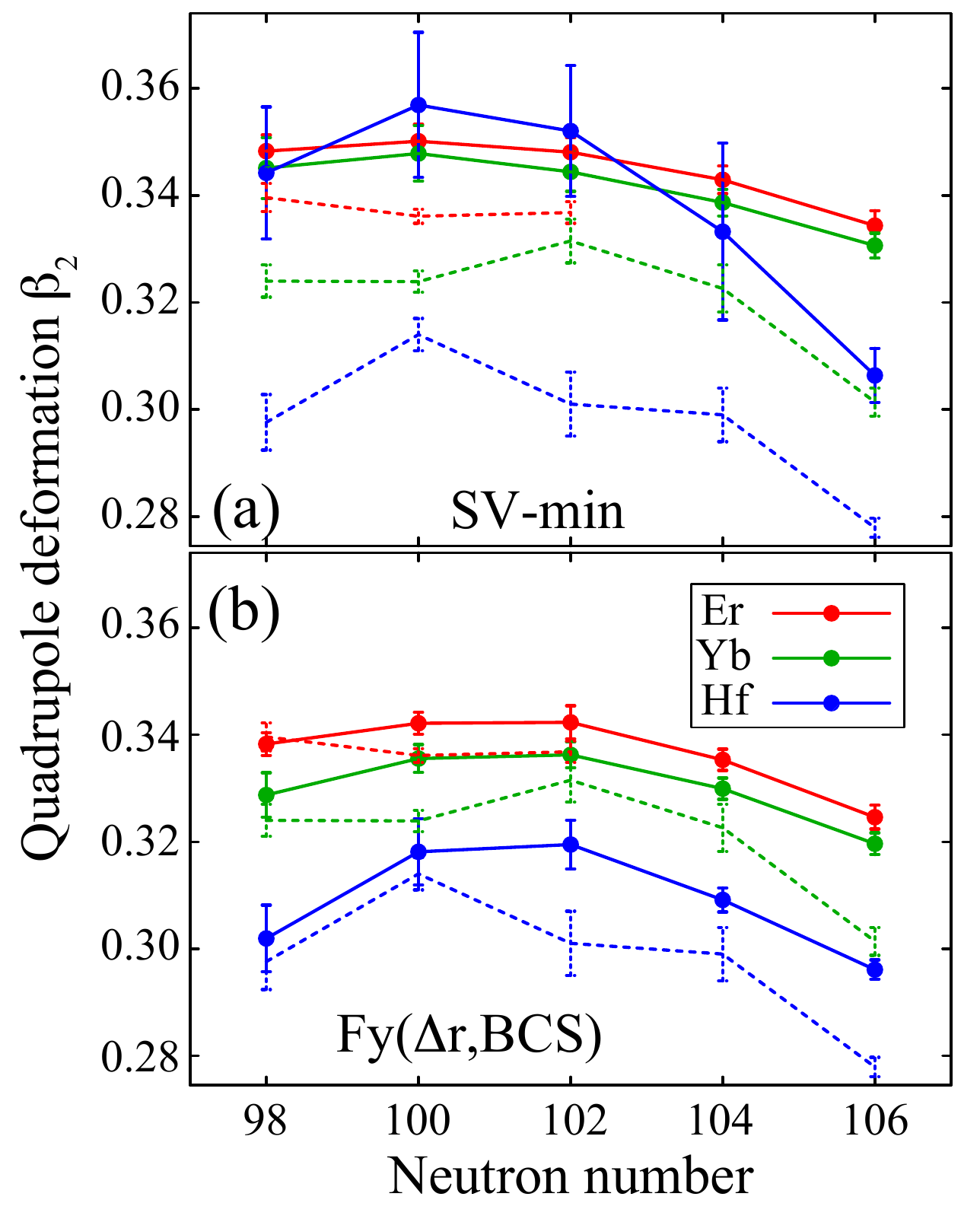}
\caption{\label{fig:beta2}
Proton quadrupole ground-state deformations $\beta_2$ for even-even Er, Yb, and Hf isotopes with $98\le N\le 106$ calculated with (a) {\SV} and (b) {\Fy}  EDFs compared to empirical values \cite{Pritychenko2016} (dashed lines). Statistical model uncertainties and experimental errors are marked.
  }
\end{figure}

\section{Results}\label{results}

The scope of this Paper is to study the structure of statistical
correlations between ground-state deformations and between charge radii in the even-even Er, Yb, and Hf isotopes  with $98\le N\le 106$. 
These nuclei lie in the center of the deformed rare-earth region \cite{Pritychenko2016}. 

The dimensionless quadrupole deformations can be  deduced from
the calculated proton quadrupole moments
\begin{equation}\label{eqbeta2}
  \beta_2
  =
  4\pi\frac{\langle r^2Y_{20}\rangle}{3ZR^2}
  \;,\;
  R=1.2A^{1/3}\,\mathrm{fm}
  \;.
\end{equation}
This quantity is directly related to the geometrical shape and thus
simplifies comparisons across different nuclei.  The average value of of the spherical radius $R$ was taken the same  as in Ref.~\cite{Pritychenko2016}.
Figure~\ref{fig:beta2} shows the calculated values of $\beta_2$ for {\SV} and {\Fy} and compares them to empirical quadrupole deformations extracted from 
the experimental transition probabilities for the lowest $2^+$ states \cite{Pritychenko2016}. All isotopes
shown in Fig.~\ref{fig:beta2} are very well deformed.
Considering the scale of Fig.~\ref{fig:beta2}, the agreement between experiment and theory is very reasonable, especially for {\Fy}.
This is not
so surprising because, as discussed in the introduction,  nuclear deformation properties are dominated by
shell topology: all reasonable nuclear models, including macroscopic-microscopic approaches   as well as various flavors of nuclear  density functional method, are bound to reproduce the
deformations of well deformed nuclei. On the other hand, appreciable model 
differences are expected  for transitional isotopes for which the concept of a rigid nuclear deformation is questionable.
\begin{figure}[!htb]
\includegraphics[width=1.0\columnwidth]{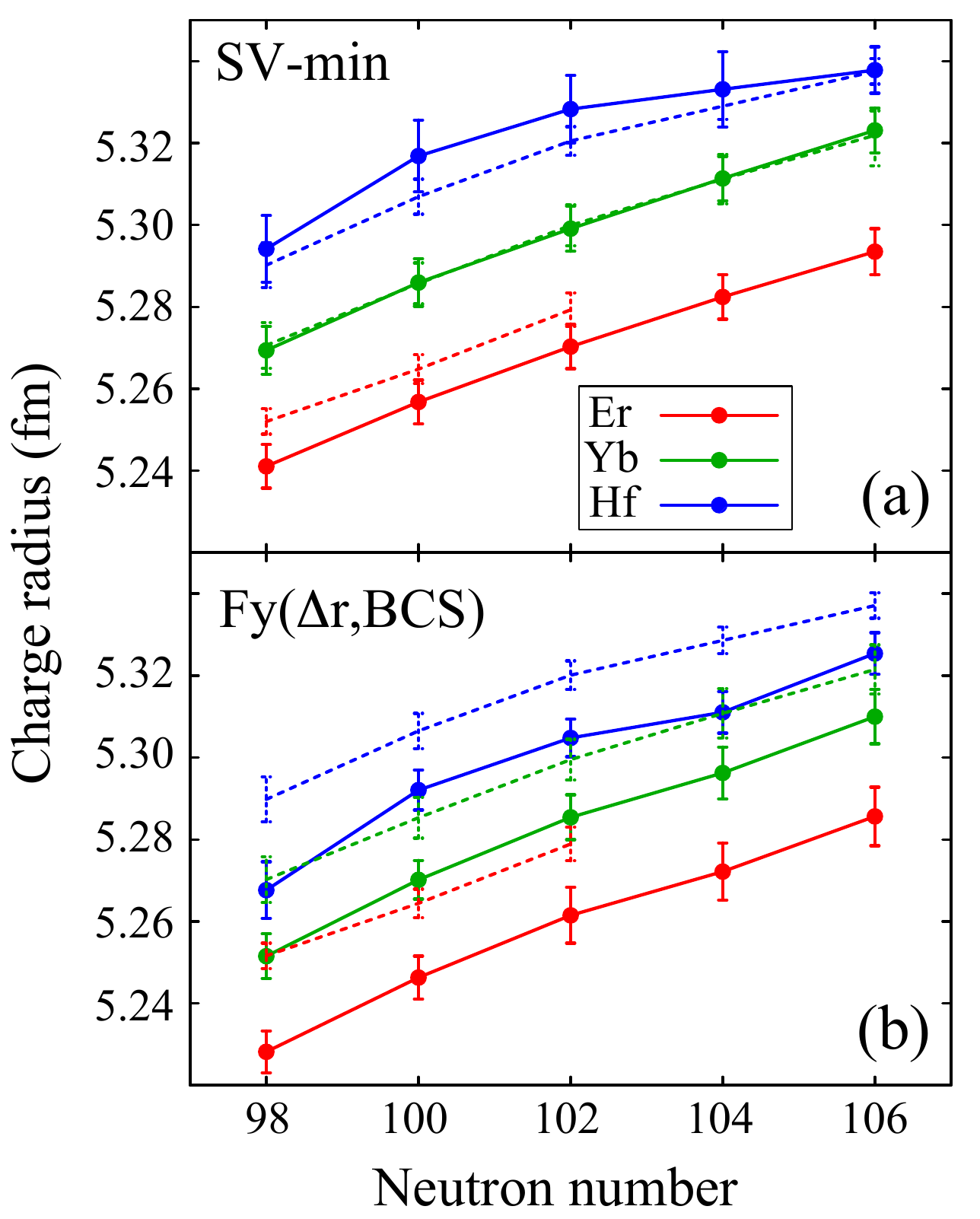}
\caption{\label{fig:rch}
Charge radii for even-even Er, Yb, and Hf isotopes with $98\le N\le 106$ calculated with (a) {\SV} and (b) {\Fy}  EDFs compared to empirical values \cite{Angeli2013} (dashed lines). Statistical model uncertainties and experimental errors are marked.
  }
\end{figure}
At the second glance, we see quantitative differences between the two
models. The values of $\beta_2$ predicted by {\SV} are 5\%-10\%  larger and the trend for the Hf isotopes differs visibly. Although the deformation is dominated
by shell structure, the final details emerge from an interplay of
Coulomb pressure, surface energy, shell effects, and pairing, which all depend on the actual model. We also note  that both models tend to predict deformations that are slightly larger than the empirical values.
But this minor mismatch is unimportant for our present study which
aims at exploring the isotopic and isotonic trends of $\beta_2$
and the statistical correlations of $\beta_2$ between the isotopes.
Coulomb pressure and surface energy change only smoothly
with $Z$ and $N$ and this should lead to strong inter-correlations. However, shell structure and pairing can fluctuate, as can already be seen from local variations of $\beta_2$ in Fig.~\ref{fig:beta2}.

The charge radii {\Rch} of the discussed Er, Yb, and Hf isotopes are displayed in Fig.~\ref{fig:rch}. The radii gradually increase with $Z$ and $N$, as expected. The fluctuations atop this  smooth behavior are seen in the differential radii and their ratios \cite{Hur2022}. The charge radii obtained in {\SV} are systematically larger than those of {\Fy}. This, together with the results for the quadrupole moments shown in Fig.~\ref{fig:beta2} suggests that the proton densities predicted by {\SV} are slightly more radially  extended.
As in the case of quadrupole deformations,  local variations of {\Rch} with $N$ and $Z$ are present.

\begin{figure}[!htb]
\includegraphics[width=1.0\columnwidth]{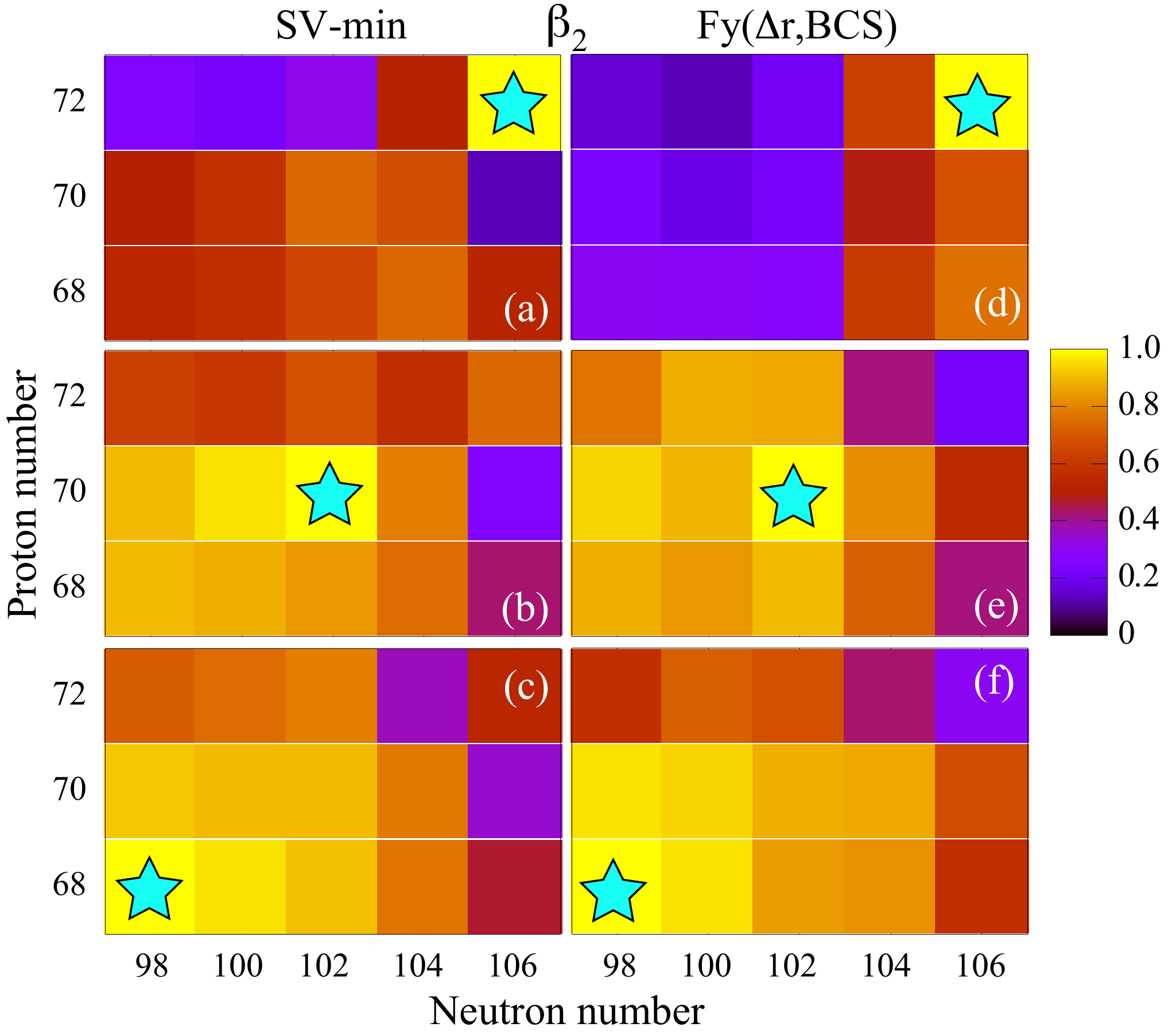}
\caption{\label{fig:beta2cod}
The CoD $R^2_{\beta_2(Z,N),\beta_2(Z',N')}$ between the quadrupole deformation
of the nucleus marked by a star and $\beta_2$ of other Er, Yb, and Hf isotopes
with  $98\le N\le 106$ calculated with (left) {\SV} and (right) {\Fy}  EDFs. 
  }
\end{figure}

\begin{figure}[!htb]
\includegraphics[width=1.0\columnwidth]{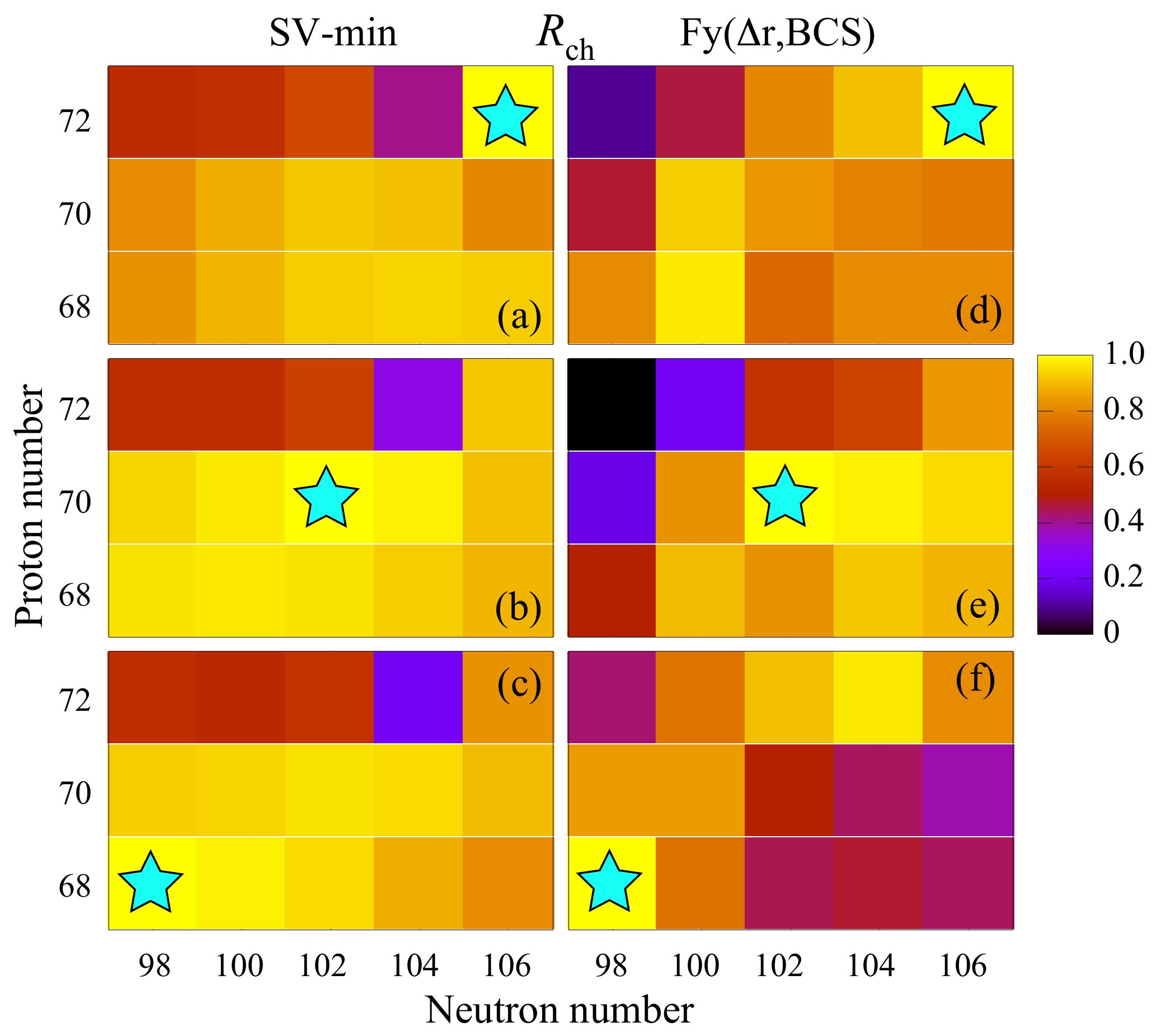}
\caption{\label{fig:rchcod}
Similar as in Fig.~\ref{fig:beta2cod} but for the charge radii.
  }
\end{figure}

\begin{figure*}[!htb]
\includegraphics[width=1.0\linewidth]{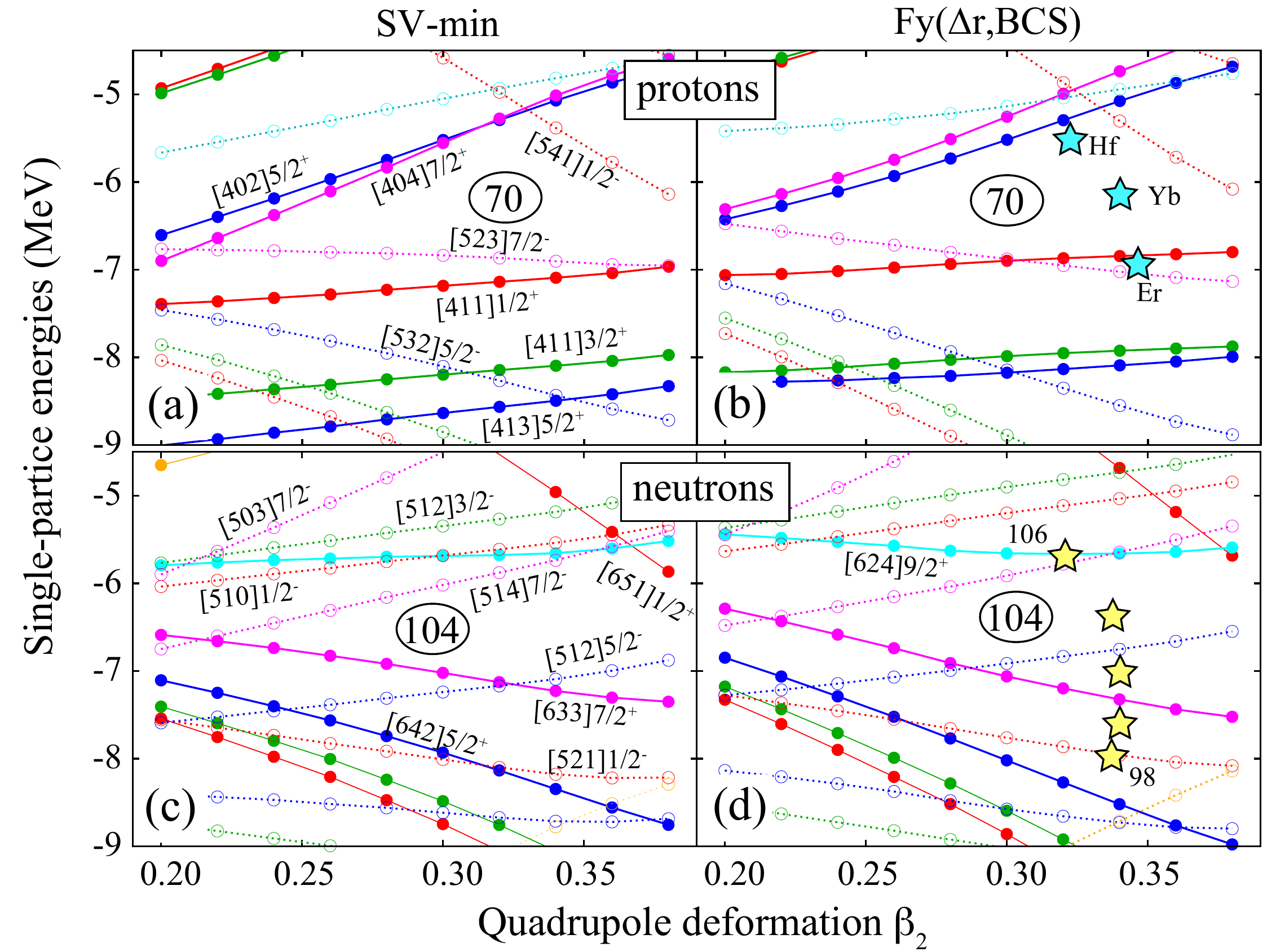}
\caption{\label{fig:splevels}
Proton (top) and neutron (bottom) single-particle energies of $^{172}$Yb calculated with  {\SV} (left)  and  {\Fy}  (right) EDFs. The asymptotic Nilsson labels $[N_{\rm osc} n_z\Lambda]\Omega^\pi$ are marked. The positions of proton Fermi levels for the $N=102$ isotones is indicated by stars in panel (b) and the neutron Fermi levels along the Yb isotopic chain - in panel (d).
  }
\end{figure*}

Figure \ref{fig:beta2cod} shows a plot of  statistical
correlations in terms of CoDs between the deformation $\beta_2$ in $^{178}$Hf
(upper panels), $^{172}$Yb (middle panels), and $^{166}$Er (lower
panels) and $\beta_2$ values of all other isotopes considered. The reference nucleus is indicated in each panel by a  star. 
Interestingly, the quadrupole deformations of  $^{172}$Yb ($N=102$) and $^{166}$Er ($N=98$) are well correlated with those of neighboring nuclei, in accordance with expectations. It is only when the neutron number approaches $N=106$ that the correlation deteriorates. The situation is different for $^{178}$Hf --- the heaviest nucleus considered. Here, the CoD values are small, even with the nearest neighbors.
The inter-nuclei correlations of charge radii are shown in Fig.~\ref{fig:rchcod}. It is seen that the values of {\Rch} are inter-correlated  better than quadrupole deformations.  But,
similar as in the $\beta_2$ case, there are regions  of surprisingly low CoDs. Particularly low correlations are predicted for  $^{176}$Hf in SV-min and $^{170}$Hf in {\Fy} for both $\beta_2$ and {\Rch}.

While the general trend is similar for both models, the quantities predicted by {\SV} are systematically better correlated than those obtained with {\Fy}. 
The significant variations of CoDs seen in Figs.~\ref{fig:beta2cod} and \ref{fig:rchcod} 
are indicative of  shell effects.
To confirm
that, we must look at the deformed shell structure in this region of nuclei.

Figure \ref{fig:splevels} shows the single particle (s.p.) energies of $^{172}$Yb as functions of $\beta_2$ (Nilsson diagram), generated by quadrupole-constrained  HF calculations. The s.p. levels are labeled by means of the asymptotic Nilsson quantum numbers $[N_{\rm osc}n_z\Lambda]\Omega^\pi$ of the stretched harmonic oscillator. The s.p. diagram of Fig.~\ref{fig:splevels} is fairly robust, i.e., it is valid for the deformed Yb region and it weakly depends on the model used (cf. Ref.~\cite{Bengtsson1989} for the modified harmonic oscillator and Woods-Saxon s.p. diagrams in this region or Ref.~\cite{Gognyplots} for Gogny-model calculations).

The proton shell structure in the deformed Yb region is defined by the pronounced deformed subshell closure at $Z=70$. At lower deformations, this gap is closed by the upsloping (oblate-driving) extruder orbitals [404]7/2$^+$ and 
[402]5/2$^+$. At larger deformations, $\beta_2>0.33$, the downsloping (prolate-driving) [541]1/2$^-$ intruder level becomes occupied at $Z=72$. Below the $Z=70$ gap, there appear two close-lying Nilsson levels: oblate-driving [411]1/2$^+$ and prolate-driving [532]7/2$^-$, which close another deformed gap at $Z=66$. These levels cross at  $\beta_2\approx 0.30$ for {\Fy} and $\beta_2\approx 0.39$ for {\SV}. 

The neutron shell structure is characterized by the deformed gap at $N=104$. This gap is closed from the above by the oblate-driving [514]7/2$^-$ and the unique-parity [624]9/2$^+$ levels, which cross at $\beta_2\approx 0.35$.  From the below, the $N=104$ gap is bounded by  the prolate-driving unique-parity [633]7/2$^+$ level and the oblate-driving [512]5/2$^-$ level, which cross at  $\beta_2\approx 0.32$ for {\SV} and $\beta_2\approx 0.28$ for {\Fy}.

\begin{figure}[!htb]
\includegraphics[width=1.0\columnwidth]{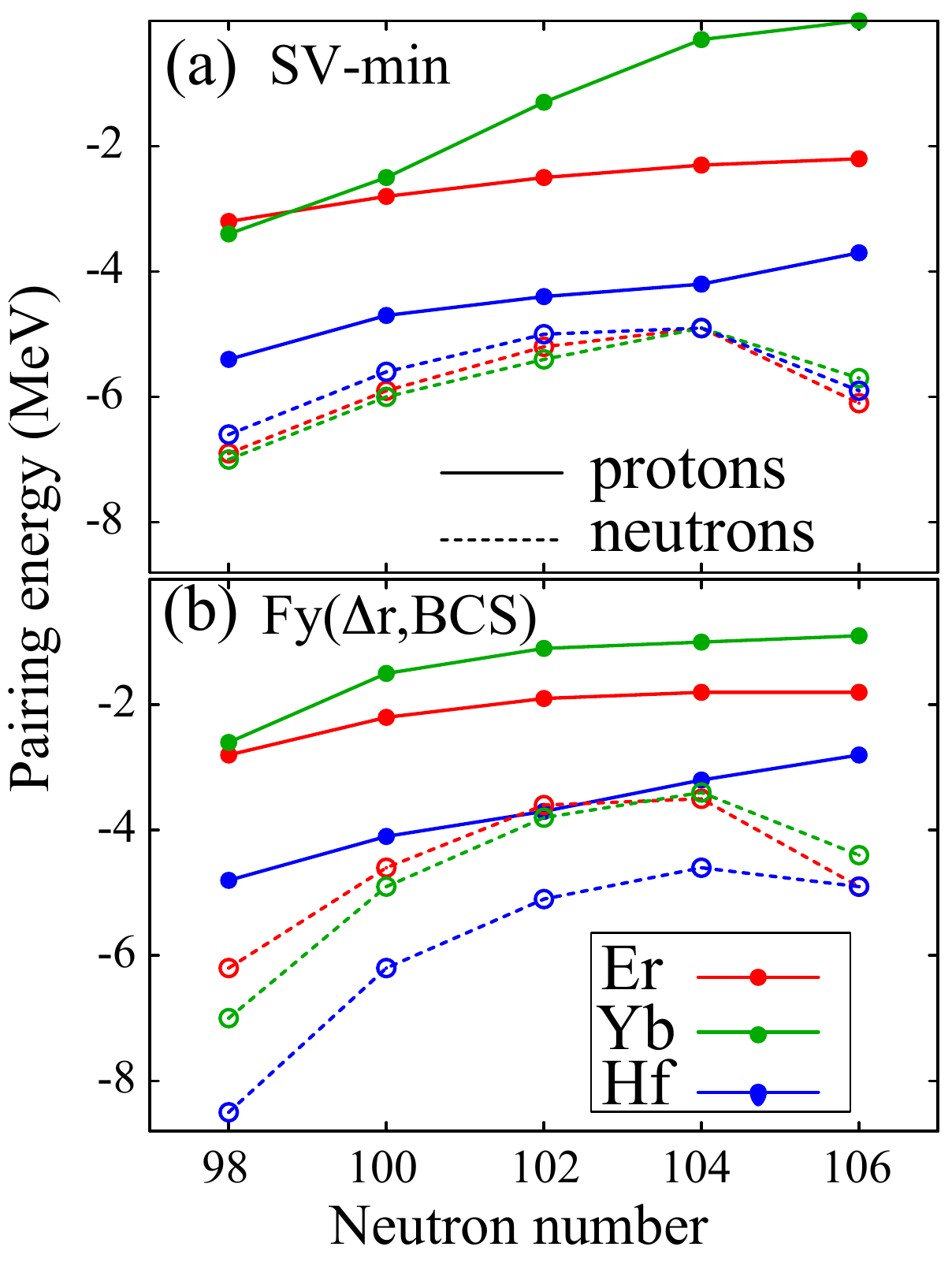}
\caption{\label{fig:epair}
Proton (solid lines) and neutron (dashed lines) pairing energies (expectation values of the pairing EDF) for even-even Er, Yb, and Hf isotopes with $98\le N\le 106$ calculated with (a) {\SV} and (b) {\Fy}  EDFs.
  }
\end{figure}

The deformed shell structure is defined by the occupations of s.p. orbits shown in Fig.~\ref{fig:splevels}. In the presence of nucleonic pairing, the s.p. occupations change gradually with particle number leading to smooth variations of nuclear observables. If pairing is weak, the transitions between intrinsic HF configurations are sharp and the underlying picture becomes diabatic. Consequently, large pairing is expected to  increase correlations between observables belonging to different nuclei. Figure~\ref{fig:epair} displays proton and neutron pairing energies of the  nuclei considered. The large deformed gap at $Z=70$ gives rise to very weak proton pairing in the Yb isotopes. The variations of neutron pairing are appreciable; they reach a minimum at the deformed neutron closure $N=104$.

The systematic trend of $\beta_2$ in Fig.~\ref{fig:beta2}
can be traced back to the s.p. diagram of Fig.~\ref{fig:splevels}. The quadrupole deformations of Er and Yb isotopes are close as the quadrupole polarization effects of [411]1/2$^+$ and [523]7/2$^-$ proton levels compensate.
The reduction of $\beta_2$ in the Hf isotopes in {\Fy} 
and  for  $N>100$ in {\Fy}
can be attributed to the occupations of the oblate-driving [404]7/2$^+$ and  [402]5/2$^+$ proton levels.
The large value of $\beta_2$ in $^{172}$Hf predicted in {\SV} is due to the filling of the $\pi$[514]1/2$^-$ level.  Finally, a reduction of $\beta_2$ when approaching $N=106$ reflects the filling of  oblate-driving [512]5/2$^-$  and [514]7/2$^-$ neutron levels. When it comes to the charge radii, the local increase of {\Rch} around $N=100, 102$ can be associated with the occupation of the neutron intruder level [633]7/2$^+$.

Having established the consistency of trends in $\beta_2$ and {\Rch} with the deformed shell-structure, let us see whether 
the details of the Nilsson plot will be reflected in the correlation systematics of Figs.~\ref{fig:beta2cod} and \ref{fig:rchcod}.
Microscopically, the energies and wave functions extruder and intruder Nilsson states (including the unique-parity states) impacting the configuration changes of the deformed Er, Yb, and Hf isotopes  are strongly influenced by the surface and spin-orbit terms of the EDFs \cite{Dobaczewski2002,Fayans2000}. Consequently, these parts of the EDFs are expected to impact statistical correlations between computed surface deformations and charge radii.

Let us begin discussion from the CoD pattern of $\beta_2$. As seen
in Figs.~\ref{fig:beta2cod}(a) and \ref{fig:beta2cod}(d),  $\beta_2$ in $^{178}$Hf is poorly correlated with quadrupole deformations of other nuclei. This nucleus is predicted to have a reduced value of $\beta_2 \approx 0.3$ compared to other systems. At this deformation,  the last two protons of $^{178}$Hf  occupy the [404]7/2$^+$ and  [402]5/2$^+$ extruder orbits, which are practically empty in the Yb and Er isotopes, as well as  in $^{170,172,174}$Hf in {\SV} in which the intruder level [541]1/2$^-$ becomes occupied at $\beta_2>0.34$. Moreover, the neutron structure of $^{178}$Hf involves the occupation of the [514]7/2$^-$ and [624]9/2$^+$ orbitals, which are empty in lighter isotopes with $N<104$.  All these configuration changes involve deformation-driving orbitals and result in reduced CoD values.

Moving on to Figs.~\ref{fig:beta2cod}(b) and \ref{fig:beta2cod}(e), the quadrupole deformation of $^{172}$Yb is correlated fairly well with the $\beta_2$ values of lighter systems. This nucleus is calculated to have $\beta_2\approx 0.34$. The decrease of correlations at $N=106$ can be associated with the filling of the neutron [624]9/2$^+$  intruder level.
The situation shown in Figs.~\ref{fig:beta2cod}(c) and \ref{fig:beta2cod}(f) for  $^{166}$Er is reminiscent of  that for $^{172}$Yb: the decrease of $\beta_2$-correlations  is seen for $N=106$ (neutron [624]9/2$^+$ occupation) and $Z=72$ (proton [541]1/2$^-$  or
[404]7/2$^+$/[402]5/2$^+$ occupation).

The pattern of inter-nuclear {\Rch}-correlations shown in Fig.~\ref{fig:rchcod} differs from that of $\beta_2$ correlations as the charge radii are expected to primarily depend on the radial properties of occupied states, hence $N_{\rm osc}$. Those are
$N_{\rm osc}=5$ proton levels [523]7/2$^-$ and [541]1/2$^-$ and
$N_{\rm osc}=6$ neutron levels [642]5/2$^+$,  [633]7/2$^+$, and
 [624]9/2$^+$.

\section{Conclusions}\label{conclusions}

In this Paper, we  investigated inter-correlations between observables in neighboring nuclei which exhibit smooth trends as a function of proton or neutron  number. To this end, we selected 15 well deformed even-even Er, Yb, and Hf isotopes in the middle of the well deformed rare earth region. The spherical shell structure in these nuclei is much fragmented by deformation effects, and the single-particle occupations are smoothed out by nucleonic pairing. 
The calculated quadrupole moments and charge radii vary gradually with $Z$ and $N$, which would intuitively suggest strong inter-correlations. To check this hypothesis, we carried out statistical correlation analysis based on covariance matrices obtained in the least-square optimization. As measure for correlations, we use the coefficient of determination  which is the square of the normalized covariance between two observables.

The calculated CoD diagrams show patterns that are surprisingly  localized as compared to the smooth  trends of observables. These local variations of CoDs reflect the underlying deformed shell structure and changes of single-particle configurations due to crossings of s.p. levels, especially high-$N_{\rm osc}$ intruder and oblate-driving extruder levels. 
In fact,  the correlation range is fairly short, i.e., it usually does not exceed $\Delta N=4$ and $\Delta Z=4$. In some extreme cases, e.g., quadrupole deformation of $^{178}$Hf in {\Fy}, the observables are hardly correlated with the values in neighboring nuclei. This finding is consistent with the results for  separation energies using  Bayesian machine learning  \cite{Neufcourt2018,Neufcourt2020a}.
Our results suggest that the frequently made assumption of strong correlations between smoothly-varying observables, which  must result in reduced statistical errors of their differences,  cannot always be justified. The recommended way to compute  statistical uncertainties on theoretical predictions and their differences, however smooth they are, remains the standard way, namely by means of covariances (or posterior distribution functions) obtained in the course of least-squares or Bayesian  calibration, see, e.g., Refs.~\cite{Klupfel2009,McDonnell2015}.

{\it Acknowledgements}.---This material is based upon work supported by the U.S.\ Department of Energy, Office of Science, Office of Nuclear Physics under award numbers DE-SC0013365 and by the National Science Foundation CSSI program under award number 2004601 (BAND collaboration).
We also thank the RRZE computing center of the Friedrich-Alexander university Erlangen/N\"urnberg for supplying resources for that work.

\bibliography{references}

\begin{thebibliography}{30}%
\makeatletter
\providecommand \@ifxundefined [1]{%
 \@ifx{#1\undefined}
}%
\providecommand \@ifnum [1]{%
 \ifnum #1\expandafter \@firstoftwo
 \else \expandafter \@secondoftwo
 \fi
}%
\providecommand \@ifx [1]{%
 \ifx #1\expandafter \@firstoftwo
 \else \expandafter \@secondoftwo
 \fi
}%
\providecommand \natexlab [1]{#1}%
\providecommand \enquote  [1]{``#1''}%
\providecommand \bibnamefont  [1]{#1}%
\providecommand \bibfnamefont [1]{#1}%
\providecommand \citenamefont [1]{#1}%
\providecommand \href@noop [0]{\@secondoftwo}%
\providecommand \href [0]{\begingroup \@sanitize@url \@href}%
\providecommand \@href[1]{\@@startlink{#1}\@@href}%
\providecommand \@@href[1]{\endgroup#1\@@endlink}%
\providecommand \@sanitize@url [0]{\catcode `\\12\catcode `\$12\catcode
  `\&12\catcode `\#12\catcode `\^12\catcode `\_12\catcode `\%12\relax}%
\providecommand \@@startlink[1]{}%
\providecommand \@@endlink[0]{}%
\providecommand \url  [0]{\begingroup\@sanitize@url \@url }%
\providecommand \@url [1]{\endgroup\@href {#1}{\urlprefix }}%
\providecommand \urlprefix  [0]{URL }%
\providecommand \Eprint [0]{\href }%
\providecommand \doibase [0]{http://dx.doi.org/}%
\providecommand \selectlanguage [0]{\@gobble}%
\providecommand \bibinfo  [0]{\@secondoftwo}%
\providecommand \bibfield  [0]{\@secondoftwo}%
\providecommand \translation [1]{[#1]}%
\providecommand \BibitemOpen [0]{}%
\providecommand \bibitemStop [0]{}%
\providecommand \bibitemNoStop [0]{.\EOS\space}%
\providecommand \EOS [0]{\spacefactor3000\relax}%
\providecommand \BibitemShut  [1]{\csname bibitem#1\endcsname}%
\let\auto@bib@innerbib\@empty
\bibitem [{\citenamefont {Reinhard}\ and\ \citenamefont
  {Otten}(1984)}]{Rei84e}%
  \BibitemOpen
  \bibfield  {author} {\bibinfo {author} {\bibfnamefont {P.-G.}\ \bibnamefont
  {Reinhard}}\ and\ \bibinfo {author} {\bibfnamefont {E.}~\bibnamefont
  {Otten}},\ }\bibfield  {title} {\enquote {\bibinfo {title} {Transition to
  deformed shapes as a nuclear {J}ahn-{T}eller effect},}\ }\href {\doibase
  10.1016/0375-9474(84)90437-8} {\bibfield  {journal} {\bibinfo  {journal}
  {Nucl. Phys. A}\ }\textbf {\bibinfo {volume} {420}},\ \bibinfo {pages} {173}
  (\bibinfo {year} {1984})}\BibitemShut {NoStop}%
\bibitem [{\citenamefont {Nazarewicz}(1993)}]{Nazarewicz1993}%
  \BibitemOpen
  \bibfield  {author} {\bibinfo {author} {\bibfnamefont {W.}~\bibnamefont
  {Nazarewicz}},\ }\bibfield  {title} {\enquote {\bibinfo {title} {Nuclear
  deformations as a spontaneous symmetry breaking},}\ }\href {\doibase
  10.1142/S0218301393000479} {\bibfield  {journal} {\bibinfo  {journal} {Int.
  J. Mod. Phys. E}\ }\textbf {\bibinfo {volume} {02}},\ \bibinfo {pages}
  {51--69} (\bibinfo {year} {1993})}\BibitemShut {NoStop}%
\bibitem [{\citenamefont {Nazarewicz}(1994)}]{Nazarewicz1994}%
  \BibitemOpen
  \bibfield  {author} {\bibinfo {author} {\bibfnamefont {W.}~\bibnamefont
  {Nazarewicz}},\ }\bibfield  {title} {\enquote {\bibinfo {title} {Microscopic
  origin of nuclear deformations},}\ }\href {\doibase
  10.1016/0375-9474(94)90037-X} {\bibfield  {journal} {\bibinfo  {journal}
  {Nucl. Phys. A}\ }\textbf {\bibinfo {volume} {574}},\ \bibinfo {pages}
  {27--49} (\bibinfo {year} {1994})}\BibitemShut {NoStop}%
\bibitem [{\citenamefont {Bertsch}(1968)}]{Bertsch1968}%
  \BibitemOpen
  \bibfield  {author} {\bibinfo {author} {\bibfnamefont {G.}~\bibnamefont
  {Bertsch}},\ }\bibfield  {title} {\enquote {\bibinfo {title} {Remark on {Y4}
  moments},}\ }\href {\doibase 10.1016/0370-2693(68)90503-0} {\bibfield
  {journal} {\bibinfo  {journal} {Phys. Lett. B}\ }\textbf {\bibinfo {volume}
  {26}},\ \bibinfo {pages} {130--131} (\bibinfo {year} {1968})}\BibitemShut
  {NoStop}%
\bibitem [{\citenamefont {Dobaczewski}\ \emph {et~al.}(1988)\citenamefont
  {Dobaczewski}, \citenamefont {Nazarewicz}, \citenamefont {Skalski},\ and\
  \citenamefont {Werner}}]{Dobaczewski1998}%
  \BibitemOpen
  \bibfield  {author} {\bibinfo {author} {\bibfnamefont {J.}~\bibnamefont
  {Dobaczewski}}, \bibinfo {author} {\bibfnamefont {W.}~\bibnamefont
  {Nazarewicz}}, \bibinfo {author} {\bibfnamefont {J.}~\bibnamefont {Skalski}},
  \ and\ \bibinfo {author} {\bibfnamefont {T.}~\bibnamefont {Werner}},\
  }\bibfield  {title} {\enquote {\bibinfo {title} {Nuclear deformation: A
  proton-neutron effect?}}\ }\href {\doibase 10.1103/PhysRevLett.60.2254}
  {\bibfield  {journal} {\bibinfo  {journal} {Phys. Rev. Lett.}\ }\textbf
  {\bibinfo {volume} {60}},\ \bibinfo {pages} {2254--2257} (\bibinfo {year}
  {1988})}\BibitemShut {NoStop}%
\bibitem [{\citenamefont {Werner}\ \emph {et~al.}(1994)\citenamefont {Werner},
  \citenamefont {Dobaczewski}, \citenamefont {Guidry}, \citenamefont
  {Nazarewicz},\ and\ \citenamefont {Sheikh}}]{Werner1994}%
  \BibitemOpen
  \bibfield  {author} {\bibinfo {author} {\bibfnamefont {T.}~\bibnamefont
  {Werner}}, \bibinfo {author} {\bibfnamefont {J.}~\bibnamefont {Dobaczewski}},
  \bibinfo {author} {\bibfnamefont {M.}~\bibnamefont {Guidry}}, \bibinfo
  {author} {\bibfnamefont {W.}~\bibnamefont {Nazarewicz}}, \ and\ \bibinfo
  {author} {\bibfnamefont {J.}~\bibnamefont {Sheikh}},\ }\bibfield  {title}
  {\enquote {\bibinfo {title} {Microscopic aspects of nuclear deformation},}\
  }\href {\doibase 10.1016/0375-9474(94)90966-0} {\bibfield  {journal}
  {\bibinfo  {journal} {Nucl. Phys. A}\ }\textbf {\bibinfo {volume} {578}},\
  \bibinfo {pages} {1 -- 30} (\bibinfo {year} {1994})}\BibitemShut {NoStop}%
\bibitem [{\citenamefont {Jänecke}(1981)}]{Janecke1981}%
  \BibitemOpen
  \bibfield  {author} {\bibinfo {author} {\bibfnamefont {J.}~\bibnamefont
  {Jänecke}},\ }\bibfield  {title} {\enquote {\bibinfo {title} {Simple
  parameterization of nuclear deformation parameters},}\ }\href {\doibase
  10.1016/0370-2693(81)90180-5} {\bibfield  {journal} {\bibinfo  {journal}
  {Phys. Lett. B}\ }\textbf {\bibinfo {volume} {103}},\ \bibinfo {pages} {1--4}
  (\bibinfo {year} {1981})}\BibitemShut {NoStop}%
\bibitem [{\citenamefont {Nazarewicz}\ and\ \citenamefont
  {Ragnarsson}(1996)}]{NazRag1996}%
  \BibitemOpen
  \bibfield  {author} {\bibinfo {author} {\bibfnamefont {W.}~\bibnamefont
  {Nazarewicz}}\ and\ \bibinfo {author} {\bibfnamefont {I.}~\bibnamefont
  {Ragnarsson}},\ }\enquote {\bibinfo {title} {Nuclear deformations},}\ in\
  \href
  {https://global.oup.com/academic/product/handbook-of-nuclear-properties-9780198517795}
  {\emph {\bibinfo {booktitle} {Handbook of Nuclear Properties}}},\ \bibinfo
  {editor} {edited by\ \bibinfo {editor} {\bibfnamefont {D.}~\bibnamefont
  {Poenaru}}\ and\ \bibinfo {editor} {\bibfnamefont {W.}~\bibnamefont
  {Greiner}}}\ (\bibinfo  {publisher} {Clarendon Press, Oxford},\ \bibinfo
  {address} {Oxford},\ \bibinfo {year} {1996})\ pp.\ \bibinfo {pages}
  {80--130}\BibitemShut {NoStop}%
\bibitem [{\citenamefont {Casten}\ \emph {et~al.}(1987)\citenamefont {Casten},
  \citenamefont {Brenner},\ and\ \citenamefont {Haustein}}]{Casten1987}%
  \BibitemOpen
  \bibfield  {author} {\bibinfo {author} {\bibfnamefont {R.~F.}\ \bibnamefont
  {Casten}}, \bibinfo {author} {\bibfnamefont {D.~S.}\ \bibnamefont {Brenner}},
  \ and\ \bibinfo {author} {\bibfnamefont {P.~E.}\ \bibnamefont {Haustein}},\
  }\bibfield  {title} {\enquote {\bibinfo {title} {Valence p-n interactions and
  the development of collectivity in heavy nuclei},}\ }\href {\doibase
  10.1103/PhysRevLett.58.658} {\bibfield  {journal} {\bibinfo  {journal} {Phys.
  Rev. Lett.}\ }\textbf {\bibinfo {volume} {58}},\ \bibinfo {pages} {658--661}
  (\bibinfo {year} {1987})}\BibitemShut {NoStop}%
\bibitem [{\citenamefont {Neufcourt}\ \emph {et~al.}(2018)\citenamefont
  {Neufcourt}, \citenamefont {Cao}, \citenamefont {Nazarewicz},\ and\
  \citenamefont {Viens}}]{Neufcourt2018}%
  \BibitemOpen
  \bibfield  {author} {\bibinfo {author} {\bibfnamefont {L.}~\bibnamefont
  {Neufcourt}}, \bibinfo {author} {\bibfnamefont {Y.}~\bibnamefont {Cao}},
  \bibinfo {author} {\bibfnamefont {W.}~\bibnamefont {Nazarewicz}}, \ and\
  \bibinfo {author} {\bibfnamefont {F.}~\bibnamefont {Viens}},\ }\bibfield
  {title} {\enquote {\bibinfo {title} {Bayesian approach to model-based
  extrapolation of nuclear observables},}\ }\href {\doibase
  10.1103/PhysRevC.98.034318} {\bibfield  {journal} {\bibinfo  {journal} {Phys.
  Rev. C}\ }\textbf {\bibinfo {volume} {98}},\ \bibinfo {pages} {034318}
  (\bibinfo {year} {2018})}\BibitemShut {NoStop}%
\bibitem [{\citenamefont {Neufcourt}\ \emph {et~al.}(2020)\citenamefont
  {Neufcourt}, \citenamefont {Cao}, \citenamefont {Giuliani}, \citenamefont
  {Nazarewicz}, \citenamefont {Olsen},\ and\ \citenamefont
  {Tarasov}}]{Neufcourt2020a}%
  \BibitemOpen
  \bibfield  {author} {\bibinfo {author} {\bibfnamefont {L.}~\bibnamefont
  {Neufcourt}}, \bibinfo {author} {\bibfnamefont {Y.}~\bibnamefont {Cao}},
  \bibinfo {author} {\bibfnamefont {S.~A.}\ \bibnamefont {Giuliani}}, \bibinfo
  {author} {\bibfnamefont {W.}~\bibnamefont {Nazarewicz}}, \bibinfo {author}
  {\bibfnamefont {E.}~\bibnamefont {Olsen}}, \ and\ \bibinfo {author}
  {\bibfnamefont {O.~B.}\ \bibnamefont {Tarasov}},\ }\bibfield  {title}
  {\enquote {\bibinfo {title} {Quantified limits of the nuclear landscape},}\
  }\href {\doibase 10.1103/PhysRevC.101.044307} {\bibfield  {journal} {\bibinfo
   {journal} {Phys. Rev. C}\ }\textbf {\bibinfo {volume} {101}},\ \bibinfo
  {pages} {044307} (\bibinfo {year} {2020})}\BibitemShut {NoStop}%
\bibitem [{\citenamefont {Bender}\ \emph {et~al.}(2003)\citenamefont {Bender},
  \citenamefont {Heenen},\ and\ \citenamefont {Reinhard}}]{Bender2003}%
  \BibitemOpen
  \bibfield  {author} {\bibinfo {author} {\bibfnamefont {M.}~\bibnamefont
  {Bender}}, \bibinfo {author} {\bibfnamefont {P.-H.}\ \bibnamefont {Heenen}},
  \ and\ \bibinfo {author} {\bibfnamefont {P.-G.}\ \bibnamefont {Reinhard}},\
  }\bibfield  {title} {\enquote {\bibinfo {title} {Self-consistent mean-field
  models for nuclear structure},}\ }\href {\doibase 10.1103/RevModPhys.75.121}
  {\bibfield  {journal} {\bibinfo  {journal} {Rev. Mod. Phys.}\ }\textbf
  {\bibinfo {volume} {75}},\ \bibinfo {pages} {121--180} (\bibinfo {year}
  {2003})}\BibitemShut {NoStop}%
\bibitem [{\citenamefont {Kl\"upfel}\ \emph {et~al.}(2009)\citenamefont
  {Kl\"upfel}, \citenamefont {Reinhard}, \citenamefont {B\"urvenich},\ and\
  \citenamefont {Maruhn}}]{Klupfel2009}%
  \BibitemOpen
  \bibfield  {author} {\bibinfo {author} {\bibfnamefont {P.}~\bibnamefont
  {Kl\"upfel}}, \bibinfo {author} {\bibfnamefont {P.-G.}\ \bibnamefont
  {Reinhard}}, \bibinfo {author} {\bibfnamefont {T.~J.}\ \bibnamefont
  {B\"urvenich}}, \ and\ \bibinfo {author} {\bibfnamefont {J.~A.}\ \bibnamefont
  {Maruhn}},\ }\bibfield  {title} {\enquote {\bibinfo {title} {Variations on a
  theme by {Skyrme}: A systematic study of adjustments of model parameters},}\
  }\href {\doibase 10.1103/PhysRevC.79.034310} {\bibfield  {journal} {\bibinfo
  {journal} {Phys. Rev. C}\ }\textbf {\bibinfo {volume} {79}},\ \bibinfo
  {pages} {034310} (\bibinfo {year} {2009})}\BibitemShut {NoStop}%
\bibitem [{\citenamefont {Reinhard}\ and\ \citenamefont
  {Nazarewicz}(2017)}]{Reinhard2017a}%
  \BibitemOpen
  \bibfield  {author} {\bibinfo {author} {\bibfnamefont {P.-G.}\ \bibnamefont
  {Reinhard}}\ and\ \bibinfo {author} {\bibfnamefont {W.}~\bibnamefont
  {Nazarewicz}},\ }\bibfield  {title} {\enquote {\bibinfo {title} {Toward a
  global description of nuclear charge radii: {Exploring the Fayans} energy
  density functional},}\ }\href {\doibase
  https://doi.org/10.1103/PhysRevC.95.064328} {\bibfield  {journal} {\bibinfo
  {journal} {Phys. Rev. C}\ }\textbf {\bibinfo {volume} {95}},\ \bibinfo
  {pages} {064328} (\bibinfo {year} {2017})}\BibitemShut {NoStop}%
\bibitem [{\citenamefont {Dobaczewski}\ \emph {et~al.}(2001)\citenamefont
  {Dobaczewski}, \citenamefont {Nazarewicz},\ and\ \citenamefont
  {Reinhard}}]{Dobaczewski2001}%
  \BibitemOpen
  \bibfield  {author} {\bibinfo {author} {\bibfnamefont {J.}~\bibnamefont
  {Dobaczewski}}, \bibinfo {author} {\bibfnamefont {W.}~\bibnamefont
  {Nazarewicz}}, \ and\ \bibinfo {author} {\bibfnamefont {P.-G.}\ \bibnamefont
  {Reinhard}},\ }\bibfield  {title} {\enquote {\bibinfo {title} {Pairing
  interaction and self-consistent densities in neutron-rich nuclei},}\ }\href
  {\doibase 10.1016/S0375-9474(01)00993-9} {\bibfield  {journal} {\bibinfo
  {journal} {Nucl. Phys. A}\ }\textbf {\bibinfo {volume} {693}},\ \bibinfo
  {pages} {361--373} (\bibinfo {year} {2001})}\BibitemShut {NoStop}%
\bibitem [{\citenamefont {Fayans}\ \emph {et~al.}(2000)\citenamefont {Fayans},
  \citenamefont {Tolokonnikov}, \citenamefont {Trykov},\ and\ \citenamefont
  {Zawischa}}]{Fayans2000}%
  \BibitemOpen
  \bibfield  {author} {\bibinfo {author} {\bibfnamefont {S.~A.}\ \bibnamefont
  {Fayans}}, \bibinfo {author} {\bibfnamefont {S.~V.}\ \bibnamefont
  {Tolokonnikov}}, \bibinfo {author} {\bibfnamefont {E.~L.}\ \bibnamefont
  {Trykov}}, \ and\ \bibinfo {author} {\bibfnamefont {D.}~\bibnamefont
  {Zawischa}},\ }\bibfield  {title} {\enquote {\bibinfo {title} {Nuclear
  isotope shifts within the local energy-density functional approach},}\ }\href
  {\doibase 10.1016/S0375-9474(00)00192-5} {\bibfield  {journal} {\bibinfo
  {journal} {Nucl. Phys. A}\ }\textbf {\bibinfo {volume} {676}},\ \bibinfo
  {pages} {49} (\bibinfo {year} {2000})}\BibitemShut {NoStop}%
\bibitem [{\citenamefont {Dobaczewski}\ \emph {et~al.}(2014)\citenamefont
  {Dobaczewski}, \citenamefont {Nazarewicz},\ and\ \citenamefont
  {Reinhard}}]{Dob14a}%
  \BibitemOpen
  \bibfield  {author} {\bibinfo {author} {\bibfnamefont {J.}~\bibnamefont
  {Dobaczewski}}, \bibinfo {author} {\bibfnamefont {W.}~\bibnamefont
  {Nazarewicz}}, \ and\ \bibinfo {author} {\bibfnamefont {P.-G.}\ \bibnamefont
  {Reinhard}},\ }\bibfield  {title} {\enquote {\bibinfo {title} {Error
  estimates of theoretical models: a guide},}\ }\href {\doibase
  10.1088/0954-3899/41/7/074001} {\bibfield  {journal} {\bibinfo  {journal} {J.
  Phys. G}\ }\textbf {\bibinfo {volume} {41}},\ \bibinfo {pages} {074001}
  (\bibinfo {year} {2014})}\BibitemShut {NoStop}%
\bibitem [{\citenamefont {Glantz}\ \emph {et~al.}(1990)\citenamefont {Glantz},
  \citenamefont {Slinker},\ and\ \citenamefont {Neilands}}]{Glantz}%
  \BibitemOpen
  \bibfield  {author} {\bibinfo {author} {\bibfnamefont {S.~A.}\ \bibnamefont
  {Glantz}}, \bibinfo {author} {\bibfnamefont {B.~K.}\ \bibnamefont {Slinker}},
  \ and\ \bibinfo {author} {\bibfnamefont {T.~B.}\ \bibnamefont {Neilands}},\
  }\href@noop {} {\emph {\bibinfo {title} {Primer of Applied Regression \&
  Analysis of Variance}}}\ (\bibinfo  {publisher} {McGraw Hill},\ \bibinfo
  {year} {1990})\BibitemShut {NoStop}%
\bibitem [{\citenamefont {Erler}\ and\ \citenamefont
  {Reinhard}(2015)}]{Erler2015}%
  \BibitemOpen
  \bibfield  {author} {\bibinfo {author} {\bibfnamefont {J.}~\bibnamefont
  {Erler}}\ and\ \bibinfo {author} {\bibfnamefont {P.-G.}\ \bibnamefont
  {Reinhard}},\ }\bibfield  {title} {\enquote {\bibinfo {title} {Error
  estimates for the {Skyrme-Hartree-Fock} model},}\ }\href {\doibase
  10.1088/0954-3899/42/3/034026} {\bibfield  {journal} {\bibinfo  {journal} {J.
  Phys. G}\ }\textbf {\bibinfo {volume} {42}},\ \bibinfo {pages} {034026}
  (\bibinfo {year} {2015})}\BibitemShut {NoStop}%
\bibitem [{\citenamefont {Reinhard}(2016)}]{Reinhard16}%
  \BibitemOpen
  \bibfield  {author} {\bibinfo {author} {\bibfnamefont {P.-G.}\ \bibnamefont
  {Reinhard}},\ }\bibfield  {title} {\enquote {\bibinfo {title} {Estimating the
  relevance of predictions from the {Skyrme–Hartree–Fock} model},}\ }\href
  {\doibase 10.1088/0031-8949/91/2/023002} {\bibfield  {journal} {\bibinfo
  {journal} {Phys. Scr.}\ }\textbf {\bibinfo {volume} {91}},\ \bibinfo {pages}
  {023002} (\bibinfo {year} {2016})}\BibitemShut {NoStop}%
\bibitem [{\citenamefont {Schuetrumpf}\ \emph {et~al.}(2017)\citenamefont
  {Schuetrumpf}, \citenamefont {Nazarewicz},\ and\ \citenamefont
  {Reinhard}}]{Schuetrumpf17}%
  \BibitemOpen
  \bibfield  {author} {\bibinfo {author} {\bibfnamefont {B.}~\bibnamefont
  {Schuetrumpf}}, \bibinfo {author} {\bibfnamefont {W.}~\bibnamefont
  {Nazarewicz}}, \ and\ \bibinfo {author} {\bibfnamefont {P.-G.}\ \bibnamefont
  {Reinhard}},\ }\bibfield  {title} {\enquote {\bibinfo {title} {Central
  depression in nucleonic densities: Trend analysis in the nuclear density
  functional theory approach},}\ }\href {\doibase 10.1103/PhysRevC.96.024306}
  {\bibfield  {journal} {\bibinfo  {journal} {Phys. Rev. C}\ }\textbf {\bibinfo
  {volume} {96}},\ \bibinfo {pages} {024306} (\bibinfo {year}
  {2017})}\BibitemShut {NoStop}%
\bibitem [{\citenamefont {Reinhard}(2018)}]{Reinhard2018d}%
  \BibitemOpen
  \bibfield  {author} {\bibinfo {author} {\bibfnamefont {P.-G.}\ \bibnamefont
  {Reinhard}},\ }\bibfield  {title} {\enquote {\bibinfo {title} {Nuclear
  density-functional theory and fission of super-heavy elements},}\ }\href
  {\doibase 10.1140/epja/i2018-12421-x} {\bibfield  {journal} {\bibinfo
  {journal} {Eur. Phys. J A}\ }\textbf {\bibinfo {volume} {54}},\ \bibinfo
  {pages} {13} (\bibinfo {year} {2018})}\BibitemShut {NoStop}%
\bibitem [{\citenamefont {Reinhard}\ \emph {et~al.}(2020)\citenamefont
  {Reinhard}, \citenamefont {Nazarewicz},\ and\ \citenamefont
  {Garcia~Ruiz}}]{Reinhard2020}%
  \BibitemOpen
  \bibfield  {author} {\bibinfo {author} {\bibfnamefont {P.-G.}\ \bibnamefont
  {Reinhard}}, \bibinfo {author} {\bibfnamefont {W.}~\bibnamefont
  {Nazarewicz}}, \ and\ \bibinfo {author} {\bibfnamefont {R.~F.}\ \bibnamefont
  {Garcia~Ruiz}},\ }\bibfield  {title} {\enquote {\bibinfo {title} {Beyond the
  charge radius: {The} information content of the fourth radial moment},}\
  }\href {\doibase 10.1103/PhysRevC.101.021301} {\bibfield  {journal} {\bibinfo
   {journal} {Phys. Rev. C}\ }\textbf {\bibinfo {volume} {101}},\ \bibinfo
  {pages} {021301(R)} (\bibinfo {year} {2020})}\BibitemShut {NoStop}%
\bibitem [{\citenamefont {Pritychenko}\ \emph {et~al.}(2016)\citenamefont
  {Pritychenko}, \citenamefont {Birch}, \citenamefont {Singh},\ and\
  \citenamefont {Horoi}}]{Pritychenko2016}%
  \BibitemOpen
  \bibfield  {author} {\bibinfo {author} {\bibfnamefont {B.}~\bibnamefont
  {Pritychenko}}, \bibinfo {author} {\bibfnamefont {M.}~\bibnamefont {Birch}},
  \bibinfo {author} {\bibfnamefont {B.}~\bibnamefont {Singh}}, \ and\ \bibinfo
  {author} {\bibfnamefont {M.}~\bibnamefont {Horoi}},\ }\bibfield  {title}
  {\enquote {\bibinfo {title} {Tables of {E2} transition probabilities from the
  first 2+ states in even–even nuclei},}\ }\href {\doibase
  10.1016/j.adt.2015.10.001} {\bibfield  {journal} {\bibinfo  {journal} {At.
  Data Nucl. Data Tables}\ }\textbf {\bibinfo {volume} {107}},\ \bibinfo
  {pages} {1–139} (\bibinfo {year} {2016})}\BibitemShut {NoStop}%
\bibitem [{\citenamefont {Angeli}\ and\ \citenamefont
  {Marinova}(2013)}]{Angeli2013}%
  \BibitemOpen
  \bibfield  {author} {\bibinfo {author} {\bibfnamefont {I.}~\bibnamefont
  {Angeli}}\ and\ \bibinfo {author} {\bibfnamefont {K.~P.}\ \bibnamefont
  {Marinova}},\ }\bibfield  {title} {\enquote {\bibinfo {title} {Table of
  experimental nuclear ground state charge radii: {An} update},}\ }\href
  {\doibase 10.1016/j.adt.2011.12.006} {\bibfield  {journal} {\bibinfo
  {journal} {At. Data Nucl. Data Tables}\ }\textbf {\bibinfo {volume} {99}},\
  \bibinfo {pages} {69--95} (\bibinfo {year} {2013})}\BibitemShut {NoStop}%
\bibitem [{\citenamefont {Hur}\ \emph {et~al.}(2022)\citenamefont {Hur},
  \citenamefont {Aude~Craik}, \citenamefont {Counts}, \citenamefont {Knyazev},
  \citenamefont {Caldwell}, \citenamefont {Leung}, \citenamefont {Pandey},
  \citenamefont {Berengut}, \citenamefont {Geddes}, \citenamefont {Nazarewicz},
  \citenamefont {Reinhard}, \citenamefont {Kawasaki}, \citenamefont {Jeon},
  \citenamefont {Jhe},\ and\ \citenamefont {Vuleti\ifmmode~\acute{c}\else
  \'{c}\fi{}}}]{Hur2022}%
  \BibitemOpen
  \bibfield  {author} {\bibinfo {author} {\bibfnamefont {J.}~\bibnamefont
  {Hur}}, \bibinfo {author} {\bibfnamefont {D.~P.~L.}\ \bibnamefont
  {Aude~Craik}}, \bibinfo {author} {\bibfnamefont {I.}~\bibnamefont {Counts}},
  \bibinfo {author} {\bibfnamefont {E.}~\bibnamefont {Knyazev}}, \bibinfo
  {author} {\bibfnamefont {L.}~\bibnamefont {Caldwell}}, \bibinfo {author}
  {\bibfnamefont {C.}~\bibnamefont {Leung}}, \bibinfo {author} {\bibfnamefont
  {S.}~\bibnamefont {Pandey}}, \bibinfo {author} {\bibfnamefont {J.~C.}\
  \bibnamefont {Berengut}}, \bibinfo {author} {\bibfnamefont {A.}~\bibnamefont
  {Geddes}}, \bibinfo {author} {\bibfnamefont {W.}~\bibnamefont {Nazarewicz}},
  \bibinfo {author} {\bibfnamefont {P.-G.}\ \bibnamefont {Reinhard}}, \bibinfo
  {author} {\bibfnamefont {A.}~\bibnamefont {Kawasaki}}, \bibinfo {author}
  {\bibfnamefont {H.}~\bibnamefont {Jeon}}, \bibinfo {author} {\bibfnamefont
  {W.}~\bibnamefont {Jhe}}, \ and\ \bibinfo {author} {\bibfnamefont
  {V.}~\bibnamefont {Vuleti\ifmmode~\acute{c}\else \'{c}\fi{}}},\ }\bibfield
  {title} {\enquote {\bibinfo {title} {Evidence of two-source {King} plot
  nonlinearity in spectroscopic search for new boson},}\ }\href {\doibase
  10.1103/PhysRevLett.128.163201} {\bibfield  {journal} {\bibinfo  {journal}
  {Phys. Rev. Lett.}\ }\textbf {\bibinfo {volume} {128}},\ \bibinfo {pages}
  {163201} (\bibinfo {year} {2022})}\BibitemShut {NoStop}%
\bibitem [{\citenamefont {Bengtsson}\ \emph {et~al.}(1989)\citenamefont
  {Bengtsson}, \citenamefont {Dudek}, \citenamefont {Nazarewicz},\ and\
  \citenamefont {Olanders}}]{Bengtsson1989}%
  \BibitemOpen
  \bibfield  {author} {\bibinfo {author} {\bibfnamefont {R.}~\bibnamefont
  {Bengtsson}}, \bibinfo {author} {\bibfnamefont {J.}~\bibnamefont {Dudek}},
  \bibinfo {author} {\bibfnamefont {W.}~\bibnamefont {Nazarewicz}}, \ and\
  \bibinfo {author} {\bibfnamefont {P.}~\bibnamefont {Olanders}},\ }\bibfield
  {title} {\enquote {\bibinfo {title} {A systematic comparison between the
  {Nilsson and Woods-Saxon} deformed shell model potentials},}\ }\href
  {\doibase 10.1088/0031-8949/39/2/002} {\bibfield  {journal} {\bibinfo
  {journal} {Phys. Scr.}\ }\textbf {\bibinfo {volume} {39}},\ \bibinfo {pages}
  {196--220} (\bibinfo {year} {1989})}\BibitemShut {NoStop}%
\bibitem [{Gog()}]{Gognyplots}%
  \BibitemOpen
  \href@noop {} {}\bibinfo {note}
  {\url{http://www-phynu.cea.fr/science_en_ligne/carte_potentiels_microscopiques/carte_potentiel_nucleaire_eng.htm}}\BibitemShut
  {NoStop}%
\bibitem [{\citenamefont {{J. Dobaczewski}}\ \emph {et~al.}(2002)\citenamefont
  {{J. Dobaczewski}}, \citenamefont {{W. Nazarewicz}},\ and\ \citenamefont
  {{M.V. Stoitsov}}}]{Dobaczewski2002}%
  \BibitemOpen
  \bibfield  {author} {\bibinfo {author} {\bibnamefont {{J. Dobaczewski}}},
  \bibinfo {author} {\bibnamefont {{W. Nazarewicz}}}, \ and\ \bibinfo {author}
  {\bibnamefont {{M.V. Stoitsov}}},\ }\bibfield  {title} {\enquote {\bibinfo
  {title} {Nuclear ground-state properties from mean-field calculations},}\
  }\href {\doibase 10.1140/epja/i2001-10218-8} {\bibfield  {journal} {\bibinfo
  {journal} {Eur. Phys. J. A}\ }\textbf {\bibinfo {volume} {15}},\ \bibinfo
  {pages} {21--26} (\bibinfo {year} {2002})}\BibitemShut {NoStop}%
\bibitem [{\citenamefont {McDonnell}\ \emph {et~al.}(2015)\citenamefont
  {McDonnell}, \citenamefont {Schunck}, \citenamefont {Higdon}, \citenamefont
  {Sarich}, \citenamefont {Wild},\ and\ \citenamefont
  {Nazarewicz}}]{McDonnell2015}%
  \BibitemOpen
  \bibfield  {author} {\bibinfo {author} {\bibfnamefont {J.~D.}\ \bibnamefont
  {McDonnell}}, \bibinfo {author} {\bibfnamefont {N.}~\bibnamefont {Schunck}},
  \bibinfo {author} {\bibfnamefont {D.}~\bibnamefont {Higdon}}, \bibinfo
  {author} {\bibfnamefont {J.}~\bibnamefont {Sarich}}, \bibinfo {author}
  {\bibfnamefont {S.~M.}\ \bibnamefont {Wild}}, \ and\ \bibinfo {author}
  {\bibfnamefont {W.}~\bibnamefont {Nazarewicz}},\ }\bibfield  {title}
  {\enquote {\bibinfo {title} {Uncertainty quantification for nuclear density
  functional theory and information content of new measurements},}\ }\href
  {\doibase 10.1103/PhysRevLett.114.122501} {\bibfield  {journal} {\bibinfo
  {journal} {Phys. Rev. Lett.}\ }\textbf {\bibinfo {volume} {114}},\ \bibinfo
  {pages} {122501} (\bibinfo {year} {2015})}\BibitemShut {NoStop}%
\end{thebibliography}%
\end{document}